\documentclass[aps,prl,twocolumn,amsmath,amssymb,superscriptaddress]{revtex4-1}
\usepackage{graphicx}
\usepackage{dcolumn}
\usepackage{bm}
\usepackage{textcomp}

\begin{document}

\title{Experimental simulation of charge conservation violation and Majorana dynamics}

\author{R. Keil$^{1,2,\ast}$, C. Noh$^{3, \ast}$, A. Rai$^{3}$, S. St\"utzer$^{1}$, S. Nolte$^{1}$, D. G. Angelakis$^{4,5}$, and A. Szameit}
\affiliation{Institute of Applied Physics, Abbe Center of Photonics, Friedrich-Schiller-Universit\"{a}t Jena, Max-Wien-Platz 1, 07743 Jena,
Germany\\
$^{2}$Institut f\"ur Experimentalphysik, Universit\"at Innsbruck, Technikerstrasse 25, 6020 Innsbruck, Austria\\
$^{3}$Centre for Quantum Technologies, National University of Singapore, 2 Science Drive 3, 117542 Singapore\\
$^{4}$School of Electronic and Computer Engineering, Technical University of Crete, Chania, Crete, 73100 Greece\\
$^{\ast}$ these authors contributed equally}


\begin{abstract}
Unphysical particles are commonly ruled out from the solution of physical equations, as they fundamentally cannot exist in any real system and,
hence, cannot be examined experimentally in a direct fashion. One of the most celebrated equations that allows unphysical solutions is the
relativistic Majorana equation\cite{Majorana} which might describe neutrinos and other exotic particles beyond the Standard Model\cite{Mohapatra}.
The equation's physical solutions, the Majorana fermions, are predicted to be their own anti-particles and as a consequence they have to be neutrally
charged\cite{Giunti}; the charged version however (called Majoranon\cite{Noh}) is, due to charge non-conservation, unphysical and cannot exist. On
the other hand, charge conservation violation has been contemplated in alternative theories associated with higher spacetime
dimensions\cite{Gninenko} or a non-vanishing photon mass\cite{Goldhaber}; theories whose exotic nature makes experimental testing with current
technology an impossible task. In our work, we present an experimental scheme based on optics with which we simulate the dynamics of a Majoranon,
involving the implementation of unphysical charge conjugation and complex conjugation. We show that the internal dynamics of the Majoranon is
fundamentally different from that of its close cousin, the Dirac particle, to illustrate the nature of the unphysical operations. For this we exploit
the fact that in quantum mechanics the wave function itself is not a measurable quantity. Therefore, wave functions of real physical particles, in
our case Dirac particles with opposite masses, can be superposed to a wave function of an unphysical particle, the Majoranon. Our results open a new
front in the field of quantum simulations of exotic phenomena, with possible applications in condensed matter physics, topological quantum computing,
and testing theories within and beyond the Standard Model with existing technology.
\end{abstract}

\maketitle

Quantum simulators were originally proposed by Richard Feynman in 1982\cite{Feynman} to tackle computational problems involving entanglement and the
superposition principle in an efficient manner. Instead of having a classical computer to enumerate quantum states (a problem that remains very often
intractable), he suggested using appropriate physical systems instead in order to reproduce the dynamics and quantum states of the problem under
study in a controllable fashion. Today, a wealth of simulation systems were successfully constructed using various architectures, such as
atoms\cite{Bloch}, trapped ions\cite{Blatt},  superconducting circuits\cite{Houck}, photons and cavity QED
set-ups\cite{Aspuru-Guzik,Hartmann,Greentree,Angelakis}. However, for most experimental implementations of simulators it is so far explicitly assumed
that the problem under consideration is actually physical and that it can be written in Hamiltonian form.

When Ettore Majorana wrote down his famous equation in 1937\cite{Majorana,Zee}, he explicitly suggested describing the characteristics of neutrinos
on its basis. He noted that Lorentz invariance not only allowed the Dirac equation, but also the expression ($\hbar\equiv c\equiv 1$)
\begin{equation}
  \label{eq:MajoranaEquation}
  i \gamma^\mu \partial_\mu \psi - m\psi_{\mathrm{c}} = 0
\end{equation}
for the wave function $\psi$ of a particle with (Majorana) mass $m$ and its charge conjugate $\psi_{\mathrm{c}}$. The appearance of the so-called
Majorana mass term points to violation of charge conservation, suggesting that a particle obeying the Majorana equation must be its own
anti-particle. For this physical reason, $\psi$ is commonly taken to be charge-neutral, i.e., the Majorana equation is frequently supplemented by the
condition $\psi = \psi_{\mathrm{c}}$ (the resulting particle is called the Majorana fermion). To date, no elementary particle has been identified as
a Majorana fermion. However, as Majorana has originally envisioned, there is the possibility that the neutrino is a Majorana fermion. In this case,
the corresponding lepton number would not be conserved and the nature of the neutrino can therefore be tested by lepton number non-conserving
processes such as neutrinoless double-beta decay\cite{Mohapatra}. The concept of Majorana fermion has also found use in condensed matter physics,
where quasiparticle excitations can be their own antiparticle. These quasi-particles, which can be found in superconducting systems for example, form
the basis for constructing non-Abelian anyons that are useful for topological quantum computation\cite{Kitaev,Nayak}.

The fact that the charged version of a Majorana fermion, the Majoranon\cite{Noh}, violates charge-conservation may provide access to physics beyond
the Standard Model. In many theories, a potential violation of charge conservation, for example associated with higher spacetime
dimensions\cite{Gninenko} or a non-vanishing photon mass\cite{Goldhaber}, is considered. In addition, simulating unphysical effects may yield
unexpected benefits in other areas, as recently shown for the case of complex conjugation that provides an efficient method to measure
entanglement\cite{DiCandia}. A simulation of the Majorana equation has been proposed for a trapped ion system\cite{Casanova}, but no experimental
data on the simulation of any unphysical particle has been reported so far, in any research field.

In this work, we break new grounds and devise an experimental scheme to simulate the dynamics of a Majoranon, thereby implementing a simulator of an
unphysical particle. To this end, we consider the Majorana equation in 1+1 dimensional spacetime, which reads for the two-component spinor $\psi =
{\psi_1 \choose \psi_2}$ as
\begin{equation}
  \label{eq:1DMajorana}
  i\partial_t\psi - \sigma_x p_x \psi + im\sigma_y \psi^* = 0 \;.
\end{equation}
Here, $p_x$ is the momentum along the spatial coordinate and we have used the representation such that
$\psi_{\mathrm{c}}=-i\sigma_z\sigma_y\psi^{\ast}$, where $\sigma_x,\sigma_y,\sigma_z$ are the Pauli matrices. One cannot directly simulate this
equation due to the fact that it contains a complex conjugation, which renders its Hamiltonian formulation impossible\cite{Casanova}. To circumvent
this problem, we exploit the fact that the field $\psi$ can be decomposed into two independent complex fields $\psi_\pm$, i.e.,
\begin{equation}
  \label{eq:psi_decomposition}
  \psi = \psi_+ + i\psi_- \; ,
\end{equation}
with $\psi_{\pm}$ being charge conjugation invariant: $-i\sigma_z\sigma_y\psi_\pm^*=\psi_\pm$\cite{Noh}. These fields thus describe charge-neutral
Majorana fermions whose single particle dynamics are described by a pair of Dirac equations, one with positive mass $m$, and one  with negative mass
$-m$:
\begin{equation}
  \label{eq:1DDirac}
  i\partial_t\psi_\pm - \sigma_x p_x \psi_\pm \mp m\sigma_z \psi_\pm = 0\; .
\end{equation}
Importantly, the Dirac equation itself is a physical equation and can be presented in Hamiltonian form. As such, it can be simulated using various
systems like trapped ions\cite{Gerritsma} or light\cite{Dreisow}. Physical operations in this decomposed Hilbert space of two independent Majorana
fermions can be used to simulate unphysical operations acting on the Majoranon such as as complex conjugation and charge conjugation to which that
evolution is intrinsically linked by Eq.~(\ref{eq:1DMajorana}). Using a photonic chip set-up we implement the proposed decomposition and simulate the
free evolution of a Majoranon. On top of demonstrating the unphysical Majoranon dynamics directly by measuring the absolute values of the spinor
components, we also compare the dynamics of a Majoranon with its Dirac \lq cousin\rq -- the same initial spinor following the Dirac evolution. Note
that discrepancies between the two arise from the difference in the term proportional to the mass that renders the Majorana equation unphysical. To
further clarify this, we evaluate the quantity $\left \langle \sigma_z \right \rangle = \sum\limits_{n} \left | \psi_{1,n} \right |^2 - \left |
\psi_{2,n} \right |^2$  to illustrate these discrepancies\cite{Lamata}. For a Dirac particle at rest ($p_x=0$, or equivalently $m \rightarrow
\infty$), it measures the population difference between the positive and negative energy branches and is a conserved quantity. On the contrary, it is
not conserved for the Majoranon at rest, but oscillates due to the unphysical mass term that continuously forces exchanges between the spinor
components. Borrowing from the physics of the Dirac particle, we will hereafter call this quantity a \lq pseudo-energy\rq ~for convenience.

\begin{figure}
\centerline{
\includegraphics[width=\columnwidth]{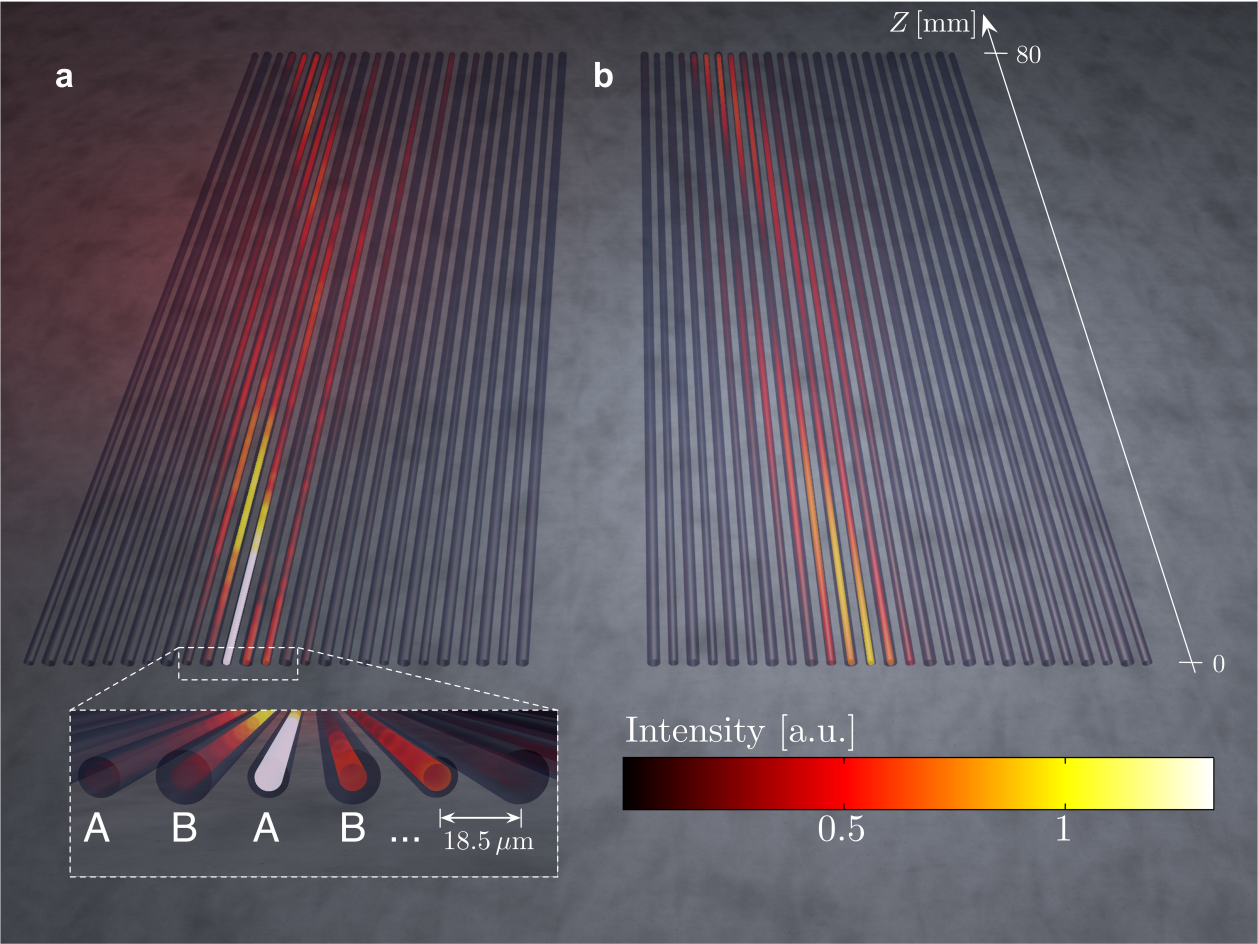}}
\caption{\label{Diracfig} Observation of photonic \textit{Zitterbewegung} in a binary waveguide array. \textbf{a,} Experimental data for a lattice of
$26$ guides. \textbf{b,} Numerical simulation using Eq. \eqref{eq:CME} with parameters $\kappa=0.064mm^{-1}$ and $\beta=0.65\kappa$ and an initial
wave packet matching the experimental conditions of the low-mass lattice (see Methods). The different refractive indices of the waveguides in the
sublattices A and B are visualised by different radii of the channels.}
\end{figure}

Our system consists of two 1+1 dimensional photonic lattices, each composed of a periodic array of waveguides that are evanescently coupled to one
another (see Methods for details on the outline of the optical simulator and waveguide fabrication). Such waveguide lattices have attracted
considerable interest and have been used in the exploration of a number of fundamental wave-transport phenomena, including Anderson
localization\cite{Schwartz}, discrete solitons\cite{Fleischer}, and photonic topological insulators\cite{Rechtsman}. In order to describe the light
evolution along the longitudinal spatial axis $Z$ in a waveguide array, one commonly employs a coupled-mode approach\cite{Yariv}, which yields
\begin{equation}
  \label{eq:CME}
  i\partial_Z \psi_k + \beta_k\psi_k + \kappa(\psi_{k+1} + \psi_{k-1}) = 0 \; ,
\end{equation}
where $\psi_k$ is the field amplitude in the $k^{\rm{th}}$ lattice site, $\kappa$ is the coupling between the waveguides, and $\beta_k$ is a position
dependent detuning. When a broad input beam with an initial phase shift of $\pi/2$ between adjacent guides is launched into a binary waveguide array
composed of two interleaved sublattices A and B with different refractive indices amounting to detunings $\pm\beta$, the light evolution can be
approximated by\cite{Longhi,Dreisow}
\begin{equation}
  \label{eq:PhotonicDirac}
  i\partial_Z\psi_{\pm} - \sigma_x \kappa p_x \psi_{\pm} \mp \beta\sigma_z \psi_{\pm} = 0\; .
\end{equation}
This is the photonic analogue of a Dirac equation for a relativistic particle with mass $\pm\beta$ (cf. Eq.~\eqref{eq:1DDirac}). The opposing signs
of the mass governing the evolution of the two spinors $\psi_{\pm}$ are implemented by an exchange of the sublattices A and B\cite{Keil}. Note that
instead of time $t$, the evolution coordinate is now the propagation distance $Z$. The beam exhibits a pronounced trembling motion around the main
trajectory, which is the photonic analogue of the famous \textit{Zitterbewegung} of a relativistic electron\cite{Schrodinger}. In our experimental
setting, we generate the desired phase distribution in the waveguide lattice by an appropriate segmentation of the waveguides (see Methods).
Figure~\ref{Diracfig}a shows an experimentally observed photonic \textit{Zitterbewegung} in a photonic lattice using such phase tailoring. A
numerical simulation of the \textit{Zitterbewegung}, based on Eq.~\eqref{eq:CME}, is shown in Fig.~\ref{Diracfig}b. The close correspondence proves
the ability to simulate the Dirac equation in a waveguide lattice.

\begin{figure}
\centerline{
\includegraphics[width=\columnwidth]{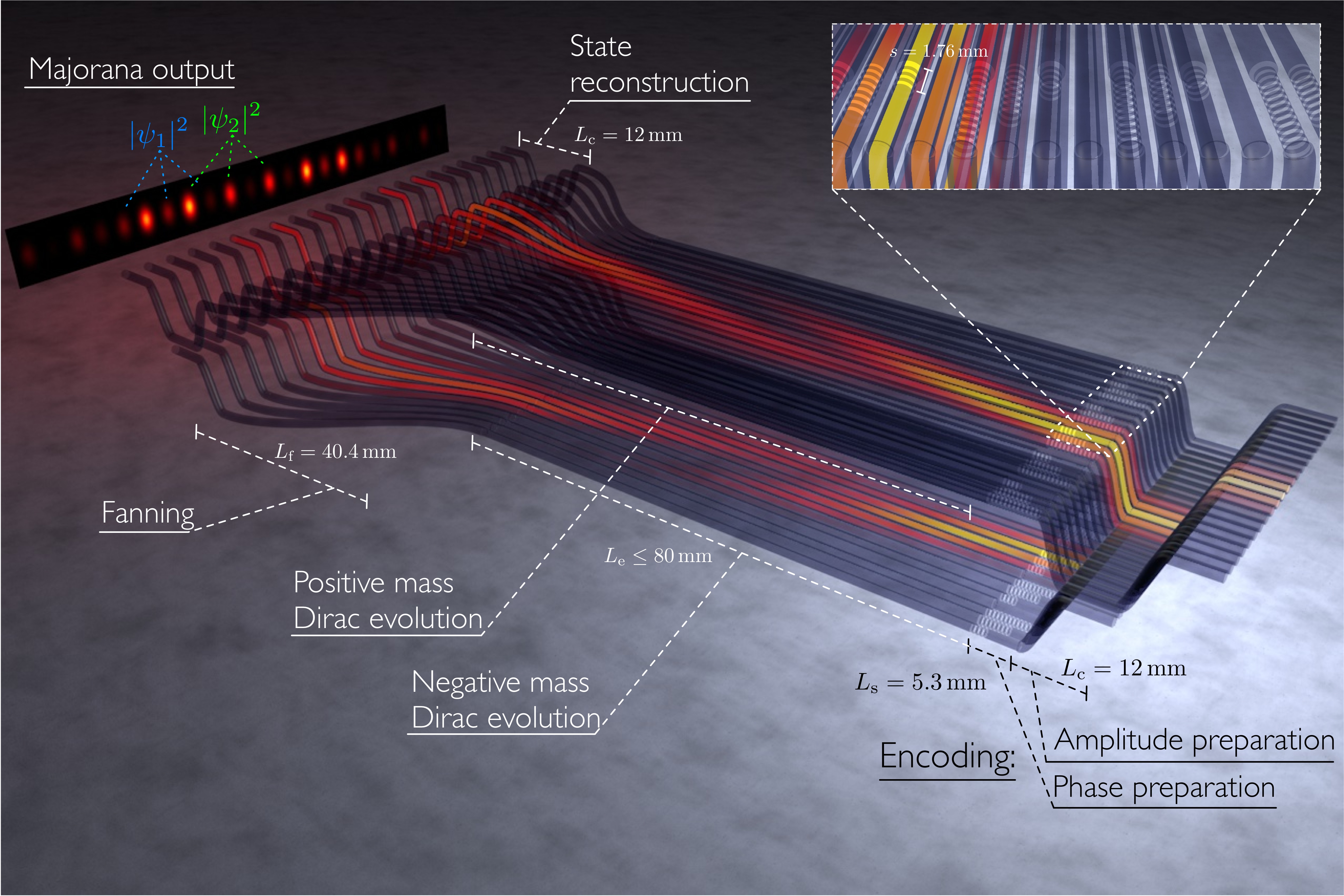}}
\caption{\label{Setup} Illustration of the waveguide sample, where two Dirac equations with opposite masses are simulated in two parallel planar
lattices. The inset shows the phase-segmentation in the upper lattice, which is used to impose a phase gradient of $\pi/2$ between adjacent guides.
The reverse segmentation profile is used in the lower plane. The calculated light intensity distribution with the same parameters as in
Fig.~\ref{Diracfig} has been superimposed onto the illustration.}
\end{figure}

\begin{figure*}
\centerline{
\includegraphics[width=\textwidth]{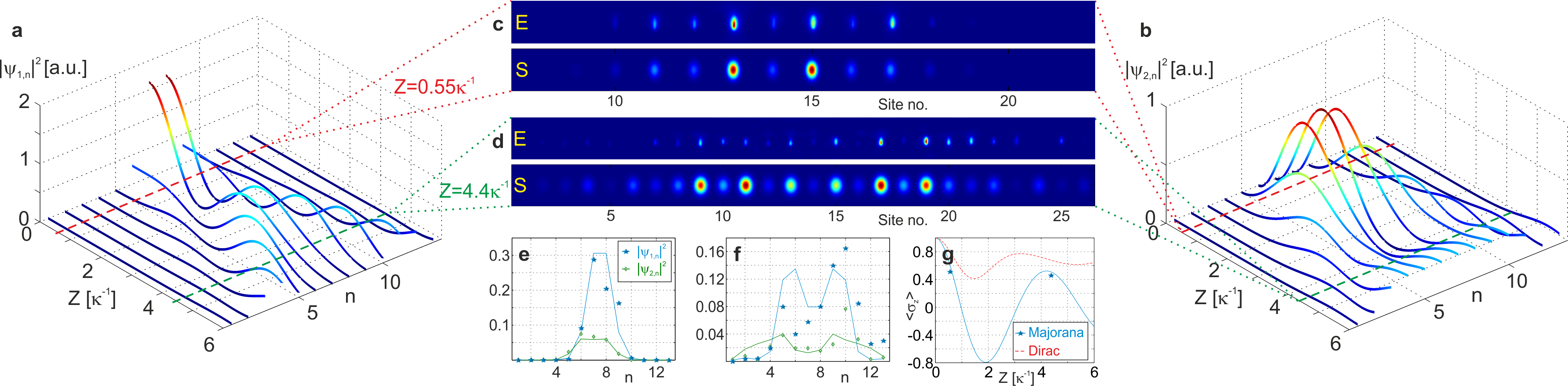}}
\caption{\label{Resultlowmass} Simulation of a Majoranon with mass $\beta=0.65\kappa$. \textbf{a,b,} Calculated intensity evolution of the first
spinor component $\psi_{1,n}$ and  the second spinor component $\psi_{2,n}$. In both panels, the number of transverse grid points $n$ and the width
of the initial wave packet correspond to the conditions in the experiment. The dashed lines indicate the evolution distances $Z$ where a measurement
is taken. \textbf{c,d,} Experimentally observed (E) and numerically simulated (S) output light intensity distributions for $Z=0.55\kappa^{-1}$ and
$Z=4.4\kappa^{-1}$. \textbf{e,} Spinor intensities reconstructed from the experimental data (symbols) in comparison to the theory (solid lines) for
the short evolution length $Z=0.55\kappa^{-1}$ and \textbf{f,} the long evolution length $Z=4.4\kappa^{-1}$. \textbf{g,} Pseudo-energy
$\left\langle\sigma_z\right\rangle$ vs $Z$. Again, the symbols represent experimental data, whereas the solid line shows the theoretical expectation.
The calculation for the corresponding Dirac spinor is shown by the dashed line.}
\end{figure*}

In our setting, we make use of exactly this fact and let two light beams propagate along two parallel planar waveguide lattices with masses of
opposite sign, such that the two Dirac equations \eqref{eq:1DDirac} are simulated in parallel, leading to \textit{Zitterbewegung} in opposite
directions (see Fig.~\ref{Setup}). After the desired propagation distance (corresponding to a specific evolution time), the amplitude distributions
are coherently combined using directional couplers between pairs of associated waveguides in the upper and lower lattices, in order to retrieve the
Majoranon wave function according to Eq. \eqref{eq:psi_decomposition} (see Methods). By construction, the first spinor component $\psi_1$ is
distributed over the odd lattice sites, whereas the second component $\psi_2$ is found on the even sites. Figure~\ref{Resultlowmass} shows our
experimental results in a system of $26$ waveguides, i.e., $n=1,\ldots,13$ discretisation points for the spinors, with a Majoranon mass
$\beta=0.65\kappa$ and $\kappa=0.064mm^{-1}$. The initial Majoranon spinor corresponds to a wavepacket with zero average momentum and $\psi_2=0$ (see
Methods). In Figs.~\ref{Resultlowmass}a and b the computed parallel evolution of both components of the Majoranon spinor is presented. We observe
that although initially all intensity is concentrated in $\psi_1$, it immediately starts to  oscillate between the two spinor components and, at the
same time, to spread along the transverse space coordinate. Using our photonic structure, we observe the population of both spinor components at two
different propagation distances. For a small effective evolution distance of $Z=L_{\rm{e,eff}}=0.55\kappa^{-1}=8.6mm$ (see Methods), the light mostly
remains in odd waveguide sites, which heralds the prevalent occupation of $\psi_1$ (Fig.~\ref{Resultlowmass}c). For a larger distance of
$Z=4.4\kappa^{-1}$, one expects another minimum of spinor 2 accompanied by extensive spreading of the wave packet (cf. Figs.~\ref{Resultlowmass}a,b).
Indeed, most of the light is again trapped in the odd channels and the entire wave packet is spread over a much larger spatial region
(Fig.~\ref{Resultlowmass}d). The individual spinor intensities, which are equivalent to the light intensities on the odd/even sites, are shown in
Figs.~\ref{Resultlowmass}e,f, together with the theoretical data. At both lengths, the population of $\psi_1$ predominates $\psi_2$.

\begin{figure*}
\centerline{
\includegraphics[width=\textwidth]{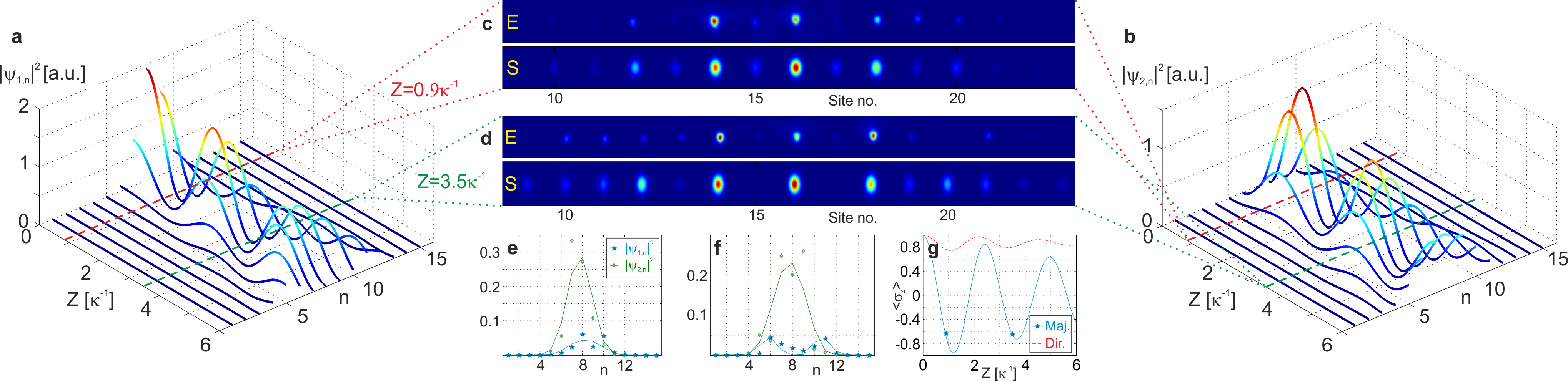}}
\caption{\label{Resulthighmass} Simulation of a Majoranon with mass $\beta=1.2\kappa$ at the two evolution distances $Z=0.9\kappa^{-1}$ and
$Z=3.5\kappa^{-1}$. The subfigures are arranged as in Fig.~\ref{Resultlowmass}.}
\end{figure*}

In Fig.~\ref{Resultlowmass}g we show the expected unphysical oscillations in the pseudo-energy of the Majoranon as discussed earlier. The measured
values of $\left \langle \sigma_z \right \rangle$ at the two evolution lengths lie in very close agreement to the expected values, while displaying
significant difference to the calculated pseudo-energy of the same initial spinor subjected to the Dirac equation~(\ref{eq:PhotonicDirac}). Note that
the oscillations in pseudo-energy for the Dirac particle and the Majoranon occur for entirely different reasons: the oscillation for the Dirac
particle occurs due to non-zero momentum components in the initial wave packet, while the oscillation for the Majoranon is mainly due to the
unphysical mass term. To elaborate on this difference further, we also study the evolution of a Majoranon for a larger mass. For this purpose, we
have implemented a second sample with a larger detuning $\beta$ between the sublattices A and B, resulting in a simulated particle mass of
$\beta=1.2\kappa$. In this system, $\kappa=0.072mm^{-1}$ and $30$ lattice sites were used. The results are summarised in Fig.~\ref{Resulthighmass}.
Due to the reduced momentum contribution in the evolution, the amplitude of the oscillation in pseudo-energy has gotten smaller for the Dirac
particle, resulting in larger discrepancies with the Majoranon, whose oscillation amplitude is not affected by the increase in mass (see
Fig.~\ref{Resulthighmass}g). The oscillation frequency, however, has increased, such that already at small distances $Z=0.9\kappa^{-1}$ mostly
$\psi_2$ is populated (see Figs.~\ref{Resulthighmass}a-c,e). After a distance of $Z=3.5\kappa^{-1}$, a further oscillation period has occurred,
leading again to a strong population of $\psi_2$. However, the transverse spreading of the wave packet is much less pronounced than for the smaller
mass of $\beta=0.65\kappa$, as clearly visible from Figs.~\ref{Resulthighmass}d,f. This is consistent with the fact that the amplitude of the
\textit{Zitterbewegung} of $\psi_{\pm}$ decreases for larger masses, whereas the frequency is increased\cite{Gerritsma,Dreisow}.

In our work, we observed the dynamics of a Majoranon wave packet which involves charge conjugation and complex conjugation. Simulating such
unphysical operations provides an entirely new approach for probing and understanding exotic phenomena and particles that cannot exist in nature,
like the Majoranon. Our approach uses the fact that even for real particles the wave function itself is not a physical entity, but only its square
modulus is. Hence, the superposition of such wave functions can result in an unphysical phenomenon, which means, conversely, that the latter can be
reproduced by simulating the individual wave functions. Many interesting questions are prompted, concerning, e.g., possible decay mechanisms of the
Majoranon, the impact of many-body effects and interactions, their scattering characteristics or possible applications in topological quantum
computing. Furthermore, we anticipate that this first demonstration of unphysical operations in the laboratory will stimulate many exciting proposals
that utilise the freedom of going beyond the `physical' operations in areas such as exotic particle physics and quantum information processing.

\begin{appendix}

\section{Methods}

\subsubsection{Design of the simulator} The experimental platform for the simulation of the Majorana equation consists of two binary waveguide
lattices, which only differ in the ordering of the two sites A and B forming a unit cell (see Fig.~\ref{Setup}). The first part of the sample is
occupied by the encoding stage (see below). In the central part, the Dirac equation \eqref{eq:PhotonicDirac} with positive (negative) mass is
simulated  over the evolution length $L_{\rm{e}}$ in the upper (lower) lattice. In this discrete setting, each spinor amplitude $\psi_{+,n}$ in unit
cell $n$ of the upper plane has its counterpart $\psi_{-,n}$ in the lower plane.

The evolution is terminated by a fan-out section of length $L_{\rm{f}}$, in which the waveguide separation is increased to some value $d$ at which no
more significant evanescent coupling takes place. This fan-out trajectory follows a harmonic curve and $L_{\rm{f}}$ is sufficiently long to ensure
that bending losses are negligibly small. Due to the gradual reduction of the coupling strength in this section, some residual evolution takes place,
which effectively extends the evolution length to some value $L_{\rm{e,eff}}>L_{\rm{e}}$\cite{Peruzzo:QuantumWalkPhotonPairs}.

Finally, all waveguide pairs of the two planes are mutually connected by vertical directional couplers of length $L_{\rm{c}}$. For balanced
couplers\cite{Yariv}, the output amplitudes in the upper ports are proportional to the discrete Majorana-spinor $\psi_n=\psi_{+,n}+i\psi_{-,n}$.
Thus, the desired recombination of the two spinors Eq. \eqref{eq:psi_decomposition} is performed in an integrated and spatially resolved fashion.

In the experiments, different evolution lengths $L_{\rm{e}}$ are used. As the total device length is fixed to $150mm$, a straight section of
identical, decoupled waveguides is introduced between the fanning and the recombination step, which preserves the field distributions. The vertical
separation of the two planes ensures an effective decoupling everywhere, except at the directional couplers.

\subsubsection{Device fabrication} Waveguides are inscribed in bulk fused silica by nonlinear absorption of focussed (numerical aperture $0.35$)
pulsed laser radiation (wavelength $800nm$, pulse duration $\tau$, pulse energy $E_{\rm{p}}$, repetition rate $100kHz$). These nonlinear absorption
processes lead to a permanent increase of the refractive index of the material. By translating the material with velocity $v_0$ on a certain path
through the focus, a waveguide channel is written\cite{Dreisow, Rechtsman, Keil}. The fabrication parameters are $\tau=150fs$, $E_{\rm{p}}=300nJ$,
$v_0=100mm/min$ for the low-mass lattice of Fig.~\ref{Resultlowmass} and $\tau=120fs$, $E_{\rm{p}}=260nJ$, $v_0=90mm/min$ for the high-mass lattice
shown in Fig.~\ref{Resulthighmass}, respectively.

The lateral waveguide separation in the evolution section is $18.5(19.5){\textrm{\textmu}m}$, for the low(high)-mass lattice and the refractive index
difference between the sublattices A and B is realised by modulating the inscription velocity by $\pm 6(14)mm/min$. The parameters of the fanning
section are $L_{\rm{f}}=40(46)mm$ and $d=40(55)\textrm{\textmu}m$. The planes are separated by $45(55)\textrm{\textmu}m$ and the couplers have a
length of $L_c=12(22)mm$, respectively.

\subsubsection{Encoding of the input state and experimental observation technique} We investigate a Majoranon wavepacket of width $\sigma$, centered
around $n_0$, with zero average momentum and occupation of only the first spinor component, i.e., $\psi_n\propto\exp(-(n-n_0)^2/2\sigma^2){1 \choose
0}$. The corresponding Dirac spinors are then given by $\psi_{+[-],n}\propto\exp(-(n-n_0)^2/2\sigma^2){1 \choose -1}\left[-{i \choose
i}\right]$.\cite{Noh} In order to ensure equal amplitude distributions in the two planes simulating $\psi_+$ and $\psi_-$, a balanced directional
coupler with a single input port is introduced at the front-end of the device, which is illuminated by a spatially extended beam in the experiment
(see Fig.~\ref{Setup}). The beam has a flat-phased Gaussian profile with a waist radius ($1/\rm{e}$-intensity) of $40(50)\textrm{\textmu}m$ for the
low(high)-mass device, corresponding to $\sigma=1.1(1.3)$, and a wavelength of $\lambda=633nm$.

Due to the mapping from Dirac spinors to light amplitudes\cite{Longhi}, the two Dirac lattices with opposing masses require a phase shift of $\pi/2$
between adjacent waveguides at the start of the evolution, but with opposite directions of the phase gradient. This is implemented by a tailored
phase segmentation of the waveguides, i.e., an intentional periodic omission of waveguide sections\cite{Keil,Szameit:Imaging}. The period of this
segmentation is $40\textrm{\textmu}m$ and the filling factor $1/2$. For $\lambda=633{nm}$, a phase delay of $j\pi/2$ is introduced by a segmented
section of length $js$, with $j=0,\ldots,3$ and $s=1.76(1.85)mm$ for the low(high)-mass lattice (see inset of Fig.~\ref{Setup}).

The intensity evolution in a single Dirac lattice is observed directly by the fluorescence of colour centres in the
waveguides\cite{Dreisow:DynamicLocalization}, whereas the evolution in the Majoranon-simulator is inferred from the measured output intensity
distributions after the recombination step.

\section{Acknowledgements}

The authors gratefully acknowledge financial support from the German Ministry of Education and Research (Center for Innovation Competence program,
grant 03Z1HN31), Thuringian Ministry for Education, Science and Culture (Research group Spacetime, grant no. 11027-514), the Singapore National
Research Foundation and Ministry of Education (partly through the Academic Research Fund Tier 3 MOE2012-T3-1-009) and the German-Israeli Foundation
for Scientific Research and Development (grant 1157-127.14/2011).

\section{Correspondence}

Correspondence and requests for materials should be addressed to D.G.A. \linebreak (email: dimitris.angelakis@qubit.org) or A.S. (email:
alexander.szameit@uni-jena.de).

\end{appendix}

\end{document}